\documentclass[a4paper,11pt]{article}
\usepackage{paper} 
\newcommand{\be}{\begin{equation}}
\newcommand{\ee}{\end{equation}}

\title{\boldmath  Rotating black hole mimicker surrounded by the string cloud}

\author[1]{Yi Yang,}
\emailAdd{yiyang@mail.gufe.edu.cn}
\affiliation[1]{School of Mathematics and Statistics, \\
Guizhou University of Finance and Economics, Guiyang, 550025, China}

\author[2]{Dong Liu,}
\emailAdd{dongliuvv@yeah.net}
\affiliation[2]{College of Physics, Guizhou University, Guiyang, 550025, China}

\author[3]{Ali \"Ovg\"un,}
\emailAdd{ali.ovgun@emu.edu.tr}
\affiliation[3]{Physics Department, Eastern Mediterranean University, Famagusta, 99628 North Cyprus via Mersin 10, Turkey}

\author[4,5]{Gaetano Lambiase,}
\emailAdd{lambiase@sa.infn.it}
\affiliation[4]{Dipartimento di Fisica ``E.R Caianiello'', Università degli Studi di Salerno, Via Giovanni Paolo II, 132 - 84084 Fisciano (SA), Italy}
\affiliation[5]{Istituto Nazionale di Fisica Nucleare - Gruppo Collegato di Salerno - Sezione di Napoli, Via Giovanni Paolo II, 132 - 84084 Fisciano (SA), Italy}

\author[2]{Zheng-Wen Long}
\emailAdd{zwlong@gzu.edu.cn }

\abstract{Traversable wormholes and regular black holes usually represent completely different scenarios. But in the black bounce spacetime they can be described by a same line element, which is very attractive. Furthermore, the black hole photos taken by EHT show that black holes have spin, so spin is an indispensable intrinsic property of black holes in the actual universe. In this work, we derive a rotating black hole mimicker surrounded by the string cloud (SC), which can be interpolated to represent regular black hole spacetime and traversable wormhole spacetime. We investigate the effect of the spin $a$ and SC parameter $L$ on the observables (shadow radius $R_s$ and distortion $\delta_s$) and energy emission rate of the black hole mimicker surrounded by the SC. We find that shadow for this spacetime is very sensitive to the $L$, i.e., the SC parameter can significantly increase the boundary of the shadow.}

\arxivnumber{}

\begin{document}
\maketitle
\flushbottom

\section{Introduction}\label{sec:intro}
Einstein proposed the extremely important general relativity (GR) in 1915. General relativity explains gravity as the curvature of spacetime. Black holes are a very important prediction of GR. In 1916, Schwarzschild \cite{Schwarzschild:1916uq}  solved the Einstein field equations to obtained a static spherical symmetric vacuum solution, which described a non-rotating black hole with only mass. In 1963, Kerr obtained a rotating, steady-state axisymmetric vacuum black hole solution \cite{Kerr:1963ud}. In the classical gravity theory, there is a singularity inside the black hole \cite{Joshi:2000fk}. Singularities represent points where geodesics terminate abruptly, therefore their appearance often heralds the collapse of the gravity theory. It is generally believed that singularities must be eliminated by quantum gravity theory effects, but so far there is no satisfactory quantum gravity theory. Therefore, studies on the properties  of classical regular black holes have attracted much attention in recent years \cite{Tsukamoto:2020bjm,Barrientos:2022avi,Ovgun:2020yuv,Ovgun:2019wej,Calza:2022ioe}.
Bardeen first obtained a kind of regular black hole without singularity, which was later called Bardeen black hole \cite{bardeen1968non}.
Simpson and Visser proposed a black bounce spacetime, which can use the same line element to describe regular black holes, one-way traversable wormholes, and two-way traversable wormholes \cite{Simpson:2018tsi}.
In Ref. \cite{Pal:2022cxb,Pal:2023wqg}, Pal et al. used the Simpson-Visser method to regularize a few naked singularity spacetimes, where they interpreted the geometry as an exact solution of non-linear electrodynamics.
Mazza et al. \cite{Mazza:2021rgq} extended the Simpson-Visser metric (SVM) to the rotating case by using the Newman-Janis (NJ) algorithm, where the NJ algorithm is a very suitable method for extending Schwarzschild-like black holes to Kerr-like black holes \cite{Newman:1965tw,Azreg-Ainou:2014nra,Azreg-Ainou:2014pra}.
Furthermore, in Ref. \cite{Shaikh:2021yux} the authors constructed the rotating version of the Simpson-Visser geometry, which can be used as an alternative to the Kerr
black hole.

In the realm of string theory, the fundamental constituents of the natural world are one-dimensional entities rather than zero-dimensional particles. When studying gravitational interactions, we can liken a collection of these strings to a one-dimensional counterpart of a dust cloud.
It is worth noting that the idea of SC was developed by Letelier in Refs. \cite{Letelier:1979ej,Letelier:1983vka}.
Taking into account the fact that, according to the string theory, the building
blocks of nature are extended (one-dimensional) strings, rather than point-like structures with zero dimensions, as assumed in quantum field theory (QFT), Letelier inferred a generalized Schwarzschild solution of Einstein's equations.
More precisely, this solution describes a spherically
symmetric cloud (dust) of strings surrounding the standard Schwarzschild black hole.
The SC alters the horizon of the black hole, compared with the Schwarzschild black hole. The entirety of this string cloud constitutes a self-contained system, ensuring the conservation of the stress-energy tensor.  In Ref. \cite{Jha:2022nzd}, the authors studied the superradiance of non-commutating rotating SV spacetime.
Recently, Rodrigues et al. \cite{Rodrigues:2022rfj} introduced a black hole mimicker surrounded by the string cloud (SC), which proves the existence of the SC makes the black bounces spacetime remain regular.
The idea of SC has been investigated in several contexts, such as missing matter (Dark Matter problem) \cite{Soleng:1993yr}, BHs in pure Lovelock gravity \cite{Toledo:2019szg}, Hawking radiation \cite{DiaseCosta:2018xyj}, AdS black holes \cite{Toledo:2019amt}, $f(R)$ gravity \cite{MoraisGraca:2017nlv}  (see, also, for example Refs. \cite{Nascimento:2023tgw,Toledo:2020xxc,MoraisGraca:2018ofn,Toledo:2018hav,MoraisGraca:2016hmv,Ghosh:2014pga,Panotopoulos:2018law,Singh:2020nwo,deMToledo:2018tjq} and references therein for further studies).

LIGO/Virgo first detected gravitational waves from the collision of two black holes in 2015 \cite{LIGOScientific:2016aoc,LIGOScientific:2016vpg,LIGOScientific:2017bnn,LIGOScientific:2016dsl}, which confirmed the existence of black holes predicted by general relativity. This opens a new era in gravitational wave astronomy. In 2019, the Event Horizon Telescope (EHT)  released the first image of a black hole, which is the shadow of the supermassive black hole M87 at the center of the Virgo elliptical galaxy \cite{EventHorizonTelescope:2019dse}. In addition, EHT released a photo of Sagittarius A* in 2022 \cite{EventHorizonTelescope:2022apq}, which is the supermassive black hole at the center of the Milky Way. The information  carried by black hole photos can help one further understand the shadow, jet and accretion process of the black hole. In the literature, there exist numerous studies focusing on the intriguing phenomenon of black hole shadows \cite{Okyay:2021nnh,Tsukamoto:2014tja,Tsukamoto:2017fxq,Shaikh:2019fpu,Shaikh:2018kfv,Kasuya:2021cpk,Kuang:2022xjp,Ovgun:2018tua,Fathi:2022ntj,Vagnozzi:2022moj,Bambi:2019tjh,Belhaj:2020okh,Belhaj:2020kwv,Islam:2022wck,KumarWalia:2022aop,Konoplya:2021slg,Zakharov:2011zz,Khodadi:2021gbc,Khodadi:2022pqh}.
In the currently released photos of black holes, the main part is the outer circular bright area and the inner dark area. The outer bright region represents the accretion disk, and the brightness of this part is not uniform. The reason for this phenomenon is that the brightness of the accretion disk is affected by the Doppler effect. The accretion disk moving toward the detector will be brighter, while moving away from the detector will be relatively darker, where the position of the bright region in the black hole shadow is closely related to the spin of a black hole. Therefore it is necessary to introduce spin into black bounces spacetime surrounded by the SC.

Motivated by EHT releases of the black hole image, our aim in this paper is to study the shadows of the rotating black hole mimicker surrounded by the SC, and analyze their characteristics. Such a study will provide some directions for the experimental detection of such a spacetime.

This work is organized as follows. In Sec. \ref{review}, we briefly review Simpson-Visser space time surrounded by the SC. In Sec. \ref{master}, the rotating black hole mimicker surrounded by the SC are derived by using the Newman-Janis method. In Sec. \ref{HJ-eq}, we obtain the geodesic equation by solving the Hamilton-Jacobi equation in the rotating black hole mimicker surrounded by the SC. In Sec. \ref{shape_shadow}, we present the shadow images of rotating black hole mimicker surrounded by the SC, and analyze the influence of spin and SC parameters on the shadow. In Sec. \ref{Rs}, we study the shadow radius, distortion, and energy emission rate of the rotating black hole mimicker surrounded by
the SC. Sec. \ref{sec:summary} is our main conclusion of this work.

\section{Review of the non-rotating black hole mimicker surrounded by the SC}\label{review}

Rodrigues et al. \cite{Rodrigues:2022rfj} derive the black bounces in a cloud of string by considering the following Einstein equations (see Appendix \ref{Appendix} for details and definitions)
\begin{equation}\label{einstein}
R_{\mu \nu}-\frac{1}{2} R g_{\mu \nu}=\kappa^2 T_{\mu \nu}=\kappa^2 T_{\mu \nu}^M+\kappa^2 T_{\mu \nu}^{SC},
\end{equation}
with
\begin{equation}\label{TensorMG}
T_{\mu \nu}^M=T_{\mu \nu}^{S V}+T_{\mu \nu}^{NMC},
\end{equation}
where $T_{\mu \nu}^{S V}$ represents the stress-energy tensor, which relates to the SVM. In addition, the information on the non-minimum coupling between the SC and the  SVM is contained in the stress-energy tensor $T_{\mu \nu}^{NMC}$. $T_{\mu \nu}^{SC}$ in Eq. (\ref{einstein}) denotes the stress-energy tensor of the SC, which reads
\begin{equation}\label{TensorCSG}
T_{\mu \nu}^{SC}=\frac{\rho \Sigma_\mu^\alpha \Sigma_{\alpha \nu}}{8 \pi \sqrt{-\gamma}},
\end{equation}
where $\rho$ denotes the density of the SC.  $T_{\mu \nu}^{SC}$ is subject to the following conservation laws $\nabla_\mu T^{SC^{\mu \nu}}$ (see Appendix \ref{Appendix}).
Rodrigues et al., according to the Einstein equations, derive the non-rotating black hole mimicker surrounded by the SC  \cite{Rodrigues:2022rfj}
\begin{equation}
d s^2=f(r) d t^2-g(r)^{-1} d r^2-\mathcal{R}^2\left(d \theta^2+\sin ^2 \theta d \phi^2\right),
\end{equation}
and
\begin{equation}\label{huar}
f(r)=g(r)=1-L-\frac{2 M}{\sqrt{\ell^2+r^2}}, \quad \mathcal{R}=\sqrt{\ell^2+r^2},
\end{equation}
where $L$ is an integration constant related to the string, and $\ell$ is a parameter responsible for central singularity regularization.
If $\ell=0$, this spacetime metric will degenerate into the Letelier spacetime, and it will degenerate into the SVM when $L=0$. This spacetime has no event horizon for $L=1$, so the SC parameter is limited to $0<L<1$.
Moreover, the remarkable feature of this spacetime is that it can change from a black hole to a wormhole, so there is a threshold in the process of this transformation
\begin{equation}
\ell_{c}=\frac{2 M}{\sqrt{1-2 L+L^2}}.
\end{equation}
As the parameter $\ell$ changes, the black bounce surrounded by the SC will represent various spacetime backgrounds: 1) when $0<\ell<\ell_c$, the spacetime is regular black hole surrounded by the SC; 2) when $\ell = \ell_c$, the spacetime is one-way wormhole surrounded by the SC; 3) When $\ell > \ell_c$, the spacetime is traversable wormhole surrounded by the SC.

\section{The rotating black hole mimicker surrounded by the SC}\label{master}
In this section, we use the NJ method to obtain the rotating black hole mimicker surrounded by the SC. In the NJ method, it is usually necessary to solve a series of differential equations to obtain the Kerr-like black hole metric \cite{Bambi:2013ufa,Xu:2021dkv,Tang:2022uwi}.
To obtain the spacetime metric of rotating black hole mimicker in a cloud of string, we first transform the spherically symmetric metric from Boyer-Lindquist to Eddington-Finkelstein coordinates. For this, we use the following coordinate transformation
\begin{equation}
\mathrm{d} u=\mathrm{d} t-\frac{\mathrm{d} r}{\sqrt{f(r) g(r)}} .
\end{equation}
After using this coordinate transformation, the spherically symmetric space-time metric becomes
\begin{equation}\label{guu}
g^{\mu \nu}=-l^\mu n^\nu-l^\nu n^\mu+m^\mu \bar{m}^\nu+m^\nu \bar{m}^\mu,
\end{equation}
where $m$ is a complex vector, $\bar{m}$ is the complex conjugate of vector $m$, and vectors $n$ and $l$ are real. Moreover, the four basis vectors ($l^\mu, n^\mu, m^\mu$ and $\bar{m}^\mu$) satisfy the modulus of the basis vector to be 1 and the orthogonality conditions between the base vectors.
For the metric of a rotating black bounce in a cloud of string, the basis vectors are
\begin{equation}
\begin{aligned}
l^\mu & =\delta_r^\mu, \\
n^\mu & =\sqrt{\frac{g(r)}{f(r)}} \delta_\mu^\mu-\frac{f(r)}{2} \delta_r^\mu, \\
m^\mu & =\frac{1}{\sqrt{2} \mathcal{R}^2} \delta_\theta^\mu+\frac{i}{\sqrt{2} r \sin \theta} \delta_\phi^\mu, \\
\bar{m}^\mu & =\frac{1}{\sqrt{2} \mathcal{R}^2} \delta_\theta^\mu-\frac{i}{\sqrt{2} r \sin \theta} \delta_\phi^\mu.
\end{aligned}
\end{equation}

The space-time coordinates between different observers satisfy the following complex transformation in the null trade
\begin{equation}
u \rightarrow u-i a \cos \theta, \quad r \rightarrow r-i a \cos \theta.
\end{equation}
When using this transformation, the metric coefficients become more complex functions of the $(r,\theta,a)$. After using this transformation, the metric functions change as follows:
$f(r) \rightarrow \mathcal{F}(r, \theta, a), g(r) \rightarrow \mathcal{G}(r, \theta, a)$ and $h(r) \rightarrow \Sigma(r, \theta, a)$. One can find that the original form can be recovered by setting $a=0$. Moreover, the null tetrad after using the complex transformation become

\begin{equation}
\begin{aligned}
l^\mu & =\delta_r^\mu, \\
n^\mu & =\sqrt{\frac{\mathcal{G}}{\mathcal{F}}} \delta_\mu^\mu-\frac{\mathcal{F}}{2} \delta_r^\mu, \\
m^\mu&=\frac{1}{\sqrt{2 \Sigma}}\left[\delta_\theta^\mu+i a \sin \theta\left(\delta_u^\mu-\delta_r^\mu\right)+\frac{i}{\sin \theta} \delta_\phi^\mu\right], \\
\bar{m}^\mu &=\frac{1}{\sqrt{2\Sigma}}\left[\delta_\theta^\mu-i a \sin \theta\left(\delta_u^\mu-\delta_r^\mu\right)-\frac{i}{\sin \theta} \delta_\phi^\mu\right] .
\end{aligned}
\end{equation}

Therefore, according to the expression of the inverse metric (\ref{guu}), we can obtain the contravariant non-zero metric components of a rotating black bounces in a cloud of string
\begin{equation}
\begin{aligned}
g^{u u} & =\frac{a^2 \sin ^2 \theta}{\Sigma}, \quad g^{u r}=-\sqrt{\frac{\mathcal{G}}{\mathcal{F}}}-\frac{a^2 \sin ^2 \theta}{\Sigma},  \\
g^{r r} & =\mathcal{F}+\frac{a^2 \sin ^2 \theta}{\Sigma}, \quad g^{r \phi}=-\frac{a}{\Sigma}, \quad g^{\theta \theta}=\frac{1}{\Sigma}, \\
g^{\phi \phi} & =\frac{1}{\Sigma \sin ^2 \theta},\quad g^{u \phi}=\frac{a}{\Sigma}.
\end{aligned}
\end{equation}

The covariant non-zero components can be read
\begin{equation}
\begin{aligned}
& g_{u u}=-\mathcal{F}, \quad g_{u r}=-\sqrt{\frac{\mathcal{F}}{\mathcal{G}}}, \quad g_{u \phi}=a\sin ^2 \theta\left(\mathcal{F}-\sqrt{\frac{\mathcal{G}}{\mathcal{F}}}\right), \\
& g_{r \phi}=a \sin ^2 \theta \sqrt{\frac{\mathcal{F}}{\mathcal{G}}} , \quad g_{\theta \theta}=\Sigma, \\
& g_{\phi \phi}=\Sigma\sin ^2 \theta+a^2\left(2 \sqrt{\frac{\mathcal{G}}{\mathcal{F}}}-\mathcal{\mathcal{F}}\right) \sin ^4 \theta.
\end{aligned}
\end{equation}
Therefore, we can obtain the rotating black bounces in a cloud of string under Eddington-Finkelstein coordinates
\begin{equation}
\begin{aligned}
d s^2= & -\mathcal{F} d u^2-2 \sqrt{\frac{\mathcal{F}}{\mathcal{G}}} d u d r+2 a \sin ^2 \theta\left(\mathcal{F}-\sqrt{\frac{\mathcal{G}}{\mathcal{F}}}\right) d u d \phi+2 a \sin ^2 \theta \sqrt{\frac{\mathcal{F}}{\mathcal{G}}} d r d \phi+\Sigma d \theta^2 \\
& +\left[\Sigma\sin ^2 \theta+a^2\left(2 \sqrt{\frac{\mathcal{G}}{\mathcal{F}}}-\mathcal{\mathcal{F}}\right) \sin ^4 \theta\right] d \phi^2.
\end{aligned}
\end{equation}
From a physical point of view, we need to transform the Eddington-Finkelstein coordinates back to Boyer-Lindquist coordinates. The corresponding coordinate transformation is
\begin{equation}
\mathrm{d} u=\mathrm{d} t+\lambda_1(r) \mathrm{d} r, \quad \mathrm{~d} \phi=\mathrm{d} \phi+\lambda_2(r) \mathrm{d} r,
\end{equation}
where the transformation functions $\lambda_1(r)$ and $\lambda_2(r)$ can use this form \cite{Azreg-Ainou:2014pra}
\begin{equation}
\lambda_1(r)=-\frac{\mathcal{K}(r)+a^2}{g(r) h(r)+a^2}, \quad \lambda_2(r)=-\frac{a}{g(r) h(r)+a^2},
\end{equation}
where
\begin{equation}
\mathcal{K}(r)=\sqrt{\frac{g(r)}{f(r)}} h(r),
\end{equation}
and
\begin{equation}
\mathcal{F}(r, \theta)=\frac{\left(g h+a^2 \cos ^2 \theta\right) \Sigma}{\left(\mathcal{K}+a^2 \cos ^2 \theta\right)^2}, \quad \mathcal{G}(r, \theta)=\frac{g h+a^2 \cos ^2 \theta}{\Sigma} .
\end{equation}
Therefore, we obtain the line element of the rotating black hole mimicker surrounded by the  string cloud
\begin{equation}\label{metric}
\begin{aligned}
\mathrm{d} s^2=&-\left[1-\frac{L(r^2+\ell^2)+2 M \sqrt{r^2+\ell^2}}{\Sigma}\right] \mathrm{d} t^2+\frac{\Sigma}{\Delta} \mathrm{d} r^2+\Sigma \mathrm{d} \theta^2\\
&-\frac{2 a\left[ 2M\sqrt{r^2+\ell^2}+L(r^2+\ell^2)\right]\sin ^2 \theta}{\Sigma} \mathrm{d} t \mathrm{~d} \phi+\frac{A \sin ^2 \theta}{\Sigma} \mathrm{d} \phi^2
\end{aligned}
\end{equation}

with
\begin{equation}
\begin{aligned}
\Sigma =&r^2+\ell^2+a^2 \cos ^2 \theta, \\
\Delta=&r^2+\ell^2+a^2- \left[2M \sqrt{r^2+\ell^2}+L(r^2+\ell^2)\right], \\
A=&\left(r^2+\ell^2+a^2\right)^2-\Delta a^2 \sin ^2 \theta .
\end{aligned}
\end{equation}
When $a=0$, the rotating black bounces in a cloud of string can degenerate into non-rotating black bounces in a cloud of string \cite{Rodrigues:2022rfj}.
When $L=0$, the rotating black bounces in a cloud of string can degenerate into the rotating black hole mimickers \cite{Mazza:2021rgq}. Moreover, the rotating black bounces in a cloud of string can degenerate into Kerr black hole metric when $L=0,\ell=0$.

\section{Geodesics around the rotating black hole mimicker surrounded by the string cloud}\label{HJ-eq}
The photons emitted from infinity will generally have two completely different endings when they pass through a black hole: the photon with a larger orbital angular momentum will be detected by an observer at infinity when it passes through some turning points; the second outcome is that photons with smaller orbital angular momentum will fall into the black hole due to the strong gravity of the black hole. We will study the geodesic equation of photons based on the spacetime metric of the rotating black hole mimicker surrounded by the SC.
To do this, we will solve the Hamilton-Jacobi equation, which can be written as
\begin{equation}
\frac{\partial S}{\partial \lambda}=-H,
\end{equation}
where $S$ represents the Jacobi action, and $\lambda$ represents the affine parameter of the geodesic. Moreover, the Hamiltonian can be given by
\begin{equation}
H=\frac{1}{2} g^{\mu \nu} \frac{\partial S}{\partial x^\mu} \frac{\partial S}{\partial x^\nu}.
\end{equation}
In the GR, the Hamilton-Jacobi equation can be read as
\begin{equation}\label{HJ}
\frac{\partial S}{\partial \lambda}=-\frac{1}{2} g^{\mu \nu} \frac{\partial S}{\partial x^\mu} \frac{\partial S}{\partial x^\nu}.
\end{equation}
To separate variables for geodesic equations, we use the separability ansatz
\begin{equation}
S=\frac{1}{2} \mu^2 \lambda-E t+\mathcal{L} \phi+S_r(r)+S_\theta(\theta)
\end{equation}
with the particle mass $\mu$. The two conserved quantities in photon motion are $E$ and $\mathcal{L}$. Moreover, $S_r(r)$ and $S_\theta(\theta)$ represent radial and angular functions, respectively. Based on these two conserved quantities and unknown functions ($S_r(r)$ and $S_\theta(\theta)$ ), we can substitute action $S$ into the Hamilton-Jacobi equation (\ref{HJ}) to obtain the following four equations
\begin{equation}
\begin{aligned}
\Delta\Sigma \frac{d t}{d \lambda} & =A- a\frac{\mathcal{L}}{E}\left[2M \sqrt{r^2+\ell^2}+L(r^2+\ell^2)\right] , \\
\Sigma \frac{d r}{d \lambda} & =\sqrt{R(r)}, \\
\Sigma \frac{d \theta}{d \lambda} & =\sqrt{\Theta(\theta)}, \\
\Sigma \frac{d \phi}{d \lambda} & =\frac{a}{\Delta}\left(E\left(r^2+\ell^2+a^2\right)-a \mathcal{L}\right)-\left(a E-\frac{\mathcal{L}}{\sin ^2 \theta}\right),
\end{aligned}
\end{equation}
where the $R(r)$ and $\Theta(\theta)$ can be read as
\begin{equation}
\begin{aligned}
R(r)&=\left[E(r^2+\ell^2+a^2)-a \mathcal{L}\right]^2-\Delta\left[\mathcal{Q}+(\mathcal{L}-aE)^2\right],\\
\Theta(\theta)&=\mathcal{Q}+a^2 E^2\cos ^2 \theta-\mathcal{L}^2 \cot ^2 \theta,
\end{aligned}
\end{equation}
where $\mathcal{Q}$ is Carter constant \cite{Carter:1968rr}, which is the integral constant for the geodesic motion of photons in a rotating black hole spacetime.
$R(r)$ and $\Theta(\theta)$ to the photon motion must be greater than 0, so there are
\begin{equation}
\frac{R(r)}{E^2}=[r^2+\ell^2+a^2-a \xi]^2-\Delta(r)\left[\eta+(\xi-a)^2\right] \geq 0,
\end{equation}
\begin{equation}
\frac{\Theta(\theta)}{E^2}=\eta+(\xi-a)^2-\left(\frac{\xi}{\sin \theta}-a \sin \theta\right)^2 \geq 0.
\end{equation}
Among them, the collision parameters determined the photon motion are defined as $\xi=\mathcal{L}/E$ and $\eta=\mathcal{Q}/E^2$. Since a black hole has an event horizon, light cannot escape from its surface, so some one may think that the size of the black hole shadow seen is determined by the black hole event horizon. In fact, there is a special surface outside the event horizon of the black hole, namely the photosphere. When a photon enters the photosphere, it will be captured by the black hole, so it cannot reach the observer at infinity. Only light rays that are outside the photosphere can reach the observer. Therefore, the boundary and size of the black hole shadow actually depend on the size of the photosphere. In general rotating black hole spacetime, the orbit of the photosphere must satisfy:  $R(r_{ph})=0,R^\prime(r_{ph})=0,R^{\prime\prime}(r_{ph})\geq0$, where $r_{ph}$ is the radius of the unstable photosphere. According to the first two conditions, we can get critical impact parameters
\begin{equation}
\xi=\frac{-\left(\left(\ell^2+r^2\right)\left(3 M+(L-1) \sqrt{\ell^2+r^2}\right)\right)+a^2\left(M+(1+L) \sqrt{\ell^2+r^2}\right)}{a\left(M+(L-1) \sqrt{\ell^2+r^2}\right)},
\end{equation}
and
\begin{equation}
\begin{aligned}
\eta=& \frac{-\left(\ell^2+r^2\right)}{a^2\left(M+(L-1) \sqrt{\ell^2+r^2}\right)^2} \\
&\times \left[\ell^4(L-1)^2+9 M^2 r^2+(L-1)^2 r^4+2 M \sqrt{\ell^2+r^2}\left(-2 a^2+3(L-1) r^2\right)\right. \\
& +\left.\ell^2\left(9 M^2+2(L-1)^2 r^2+6(L-1) M \sqrt{\ell^2+r^2}\right)\right].
\end{aligned}
\end{equation}
\section{The shadow of rotating black hole mimicker surrounded by the SC}\label{shape_shadow}
In the previous section we calculate the geodesics of photon motion. Based on the geodesics we can study the photon motion in the rotating black bounces surrounded by the SC measured by the observer. To facilitate the calculation of shadow images, the general approach is to use the celestial coordinates $\alpha$ and $\beta$. This coordinate is on the celestial plane perpendicular to the line connecting the observer to the center of spacetime. For an observer at ($r_0,\theta_0$), the celestial coordinates can be written as \cite{Hioki:2009na}
\begin{equation}
\alpha=\lim _{r_0 \rightarrow \infty}\left(-r_0^2 \sin \theta_0 \frac{d \phi}{d r}\right),
\end{equation}
and
\begin{equation}
\beta=\lim _{r_0 \rightarrow \infty}\left(r_0^2 \frac{d \theta}{d r}\right),
\end{equation}
where $r_0$ is the distance from the observer to the black hole, and $\theta_0$ is the angle between the line connecting the observer to the center of the black hole and the rotation axis of the black hole.
For asymptotically flat spacetime metrics, the celestial coordinates can be reduced to
\begin{equation}
\alpha=-\frac{\xi}{\sin \theta_0},
\end{equation}
and
\begin{equation}
\beta= \pm \sqrt{\eta+a^2 \cos ^2 \theta_0-\xi^2 \cot ^2 \theta_0}.
\end{equation}
For our rotating black hole mimicker surrounded by the SC, when $\theta_0= \pi/2$, the celestial coordinates can be written as
\begin{equation}\label{alpha32}
\alpha=- \frac{a^2\left(M+(1+L) \sqrt{\ell^2+r^2}\right)-\left(\left(\ell^2+r^2\right)\left(3 M+(L-1) \sqrt{\ell^2+r^2}\right)\right)}{a\left(M+(L-1) \sqrt{\ell^2+r^2}\right)},
\end{equation}
and

\begin{equation}\label{beta33}
\begin{aligned}
\beta=&\pm \left\{-\frac{\left(\ell^2+r^2\right)}{a^2\left(M+(L-1) \sqrt{\ell^2+r^2}\right)^2}\right.\left[\ell^4(L-1)^2+9 M^2 r^2+(L-1)^2 r^4+\right. \\
& \left.\left.2 M \sqrt{\ell^2+r^2}\left(3(L-1) r^2-2 a^2\right)+\ell^2\left(9 M^2+2(L-1)^2 r^2+6(L-1) M \sqrt{\ell^2+r^2}\right)\right]\right\}^{1/2}.
\end{aligned}
\end{equation}

Now we can use equations (\ref{alpha32}) and (\ref{beta33}) to calculate the shadow of rotating metric surrounded by the SC, so that we can study the influence of the SC on the rotating black hole mimicker.
\begin{figure}[b!]
\hspace{-0.6cm}
\includegraphics[scale=0.4]{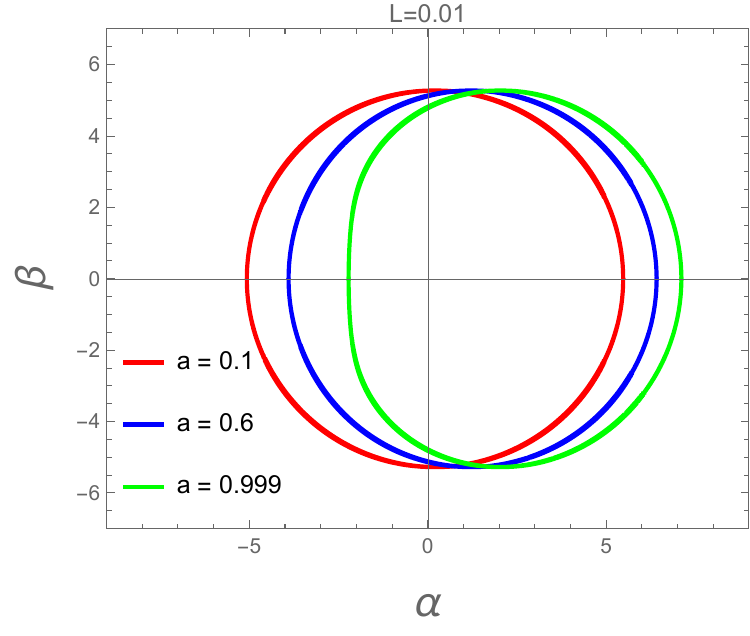}
\includegraphics[scale=0.4]{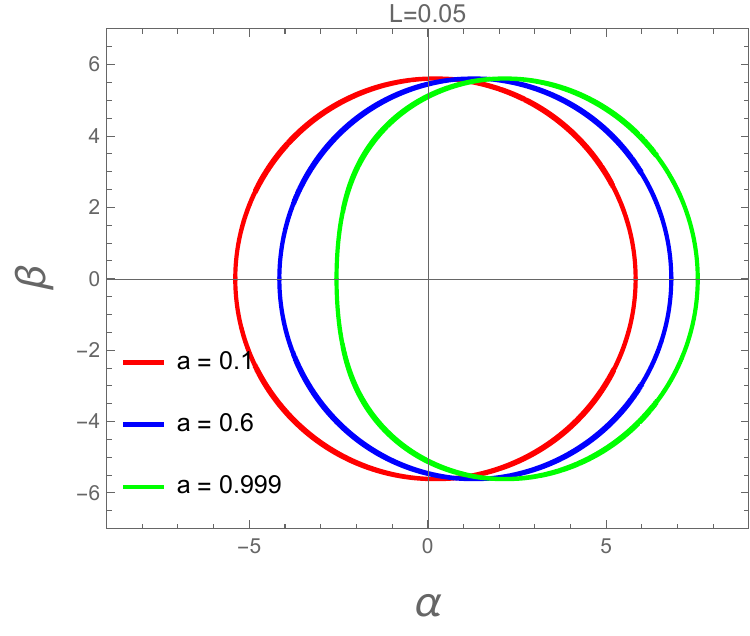}
\includegraphics[scale=0.4]{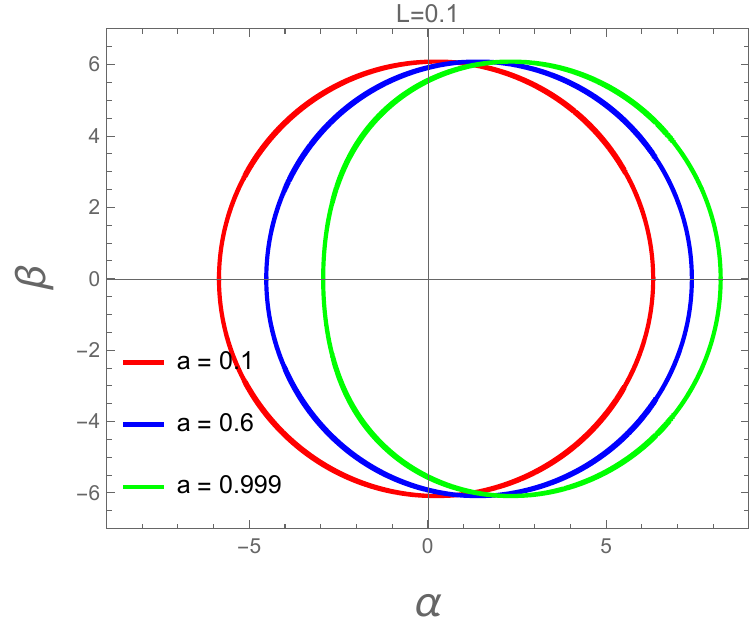}
\setlength{\abovecaptionskip}{-0.2cm}
\setlength{\belowcaptionskip}{0.4cm}
\caption{The shape of the shadow of the rotating black hole mimicker surrounded by the SC for different $a$ with $M=1,\ell=0.1$. The values of $L$ from left to right are 0.01, 0.05, 0.1, respectively.}
\label{a_l_vary}
\hspace{-0.6cm}
\includegraphics[scale=0.4]{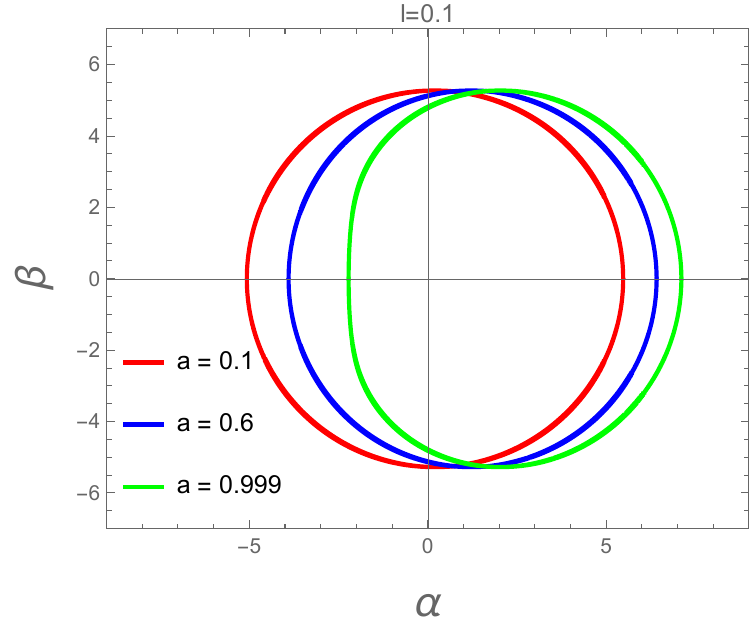}
\includegraphics[scale=0.4]{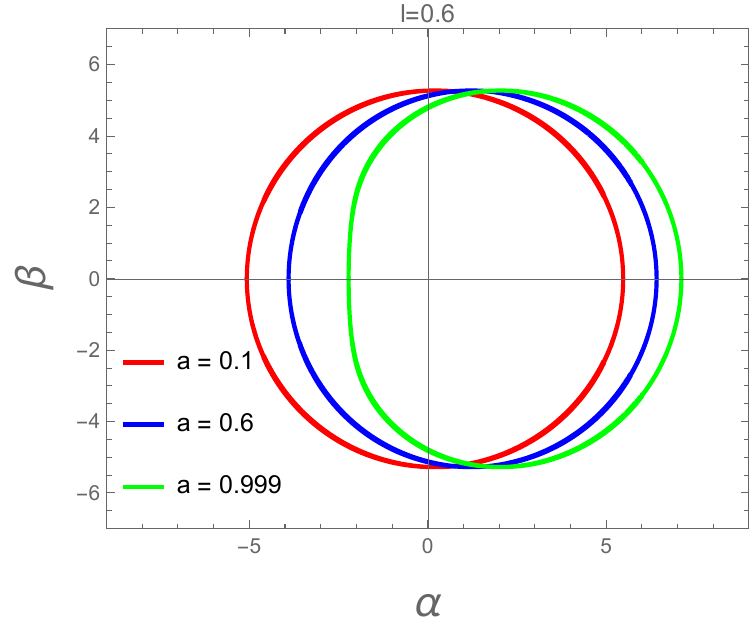}
\includegraphics[scale=0.4]{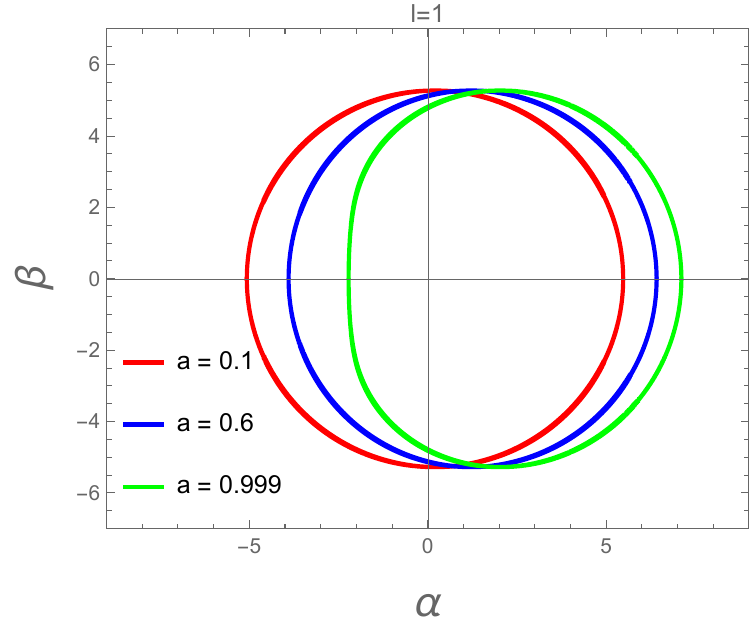}
\setlength{\abovecaptionskip}{-0.2cm}
\setlength{\belowcaptionskip}{0.4cm}
\caption{The shape of the shadow of the rotating black hole mimicker surrounded by the SC for different $a$ with $M=1,L=0.01$. The values of $\ell$ from left to right are 0.1, 0.6, 1, respectively.}
\label{a_L_vary}
\hspace{-0.6cm}
\includegraphics[scale=0.4]{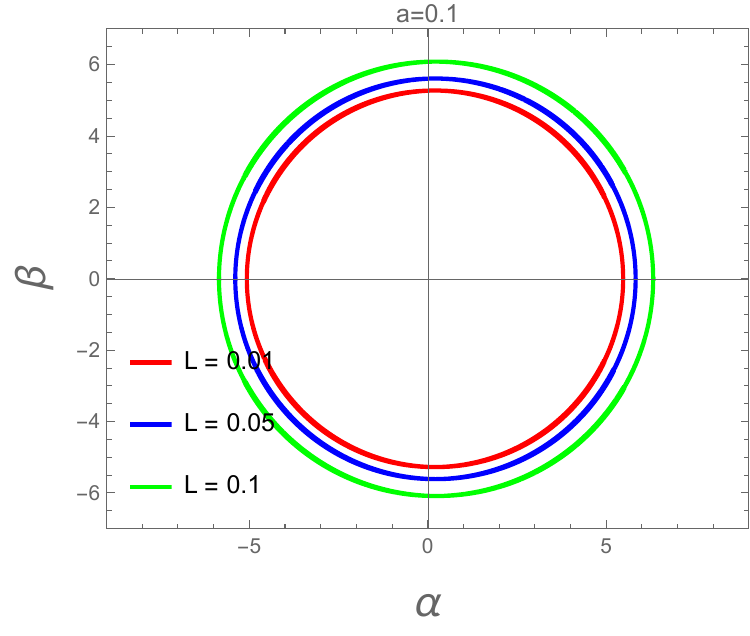}
\includegraphics[scale=0.4]{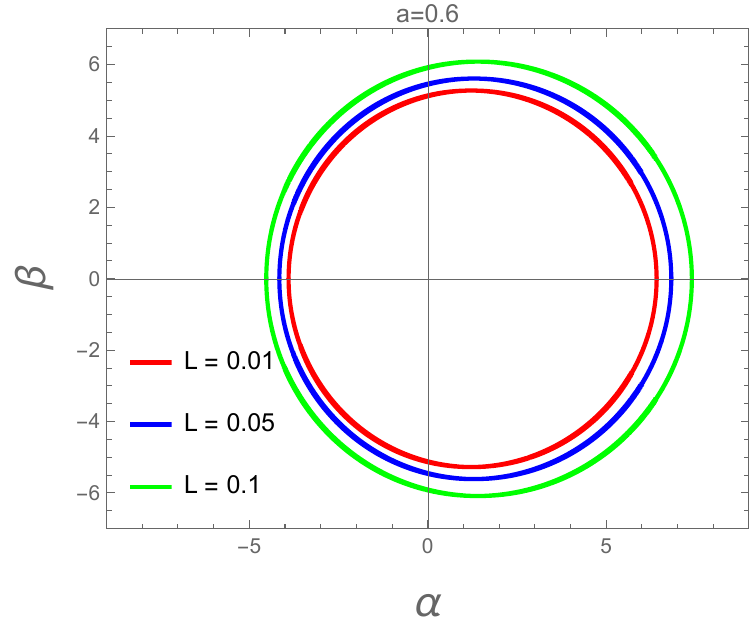}
\includegraphics[scale=0.4]{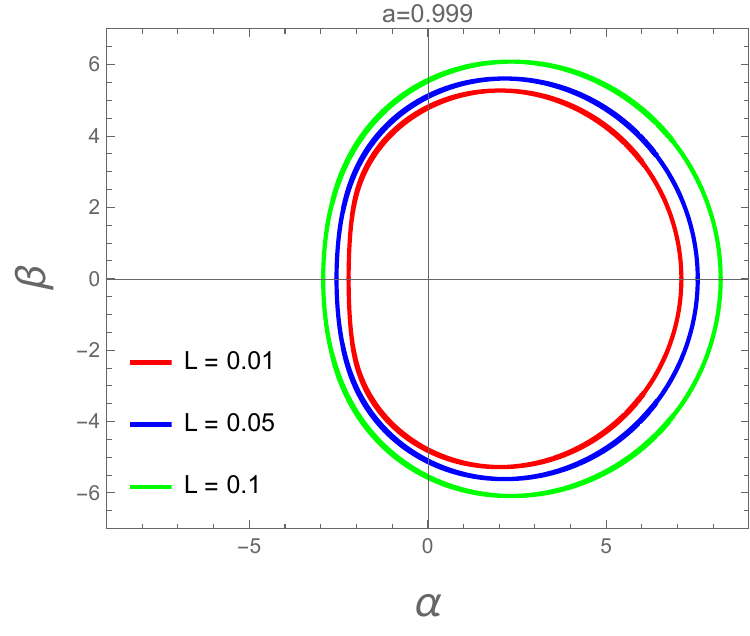}
\setlength{\abovecaptionskip}{-0.2cm}
\setlength{\belowcaptionskip}{0.4cm}
\caption{The shape of the shadow of the rotating black hole mimicker surrounded by the SC for different $L$ with $M=1,\ell=0.1$. The values of $a$ from left to right are 0.1, 0.6, 0.999, respectively. }
\label{L_vary}
\hspace{-0.6cm}
\includegraphics[scale=0.4]{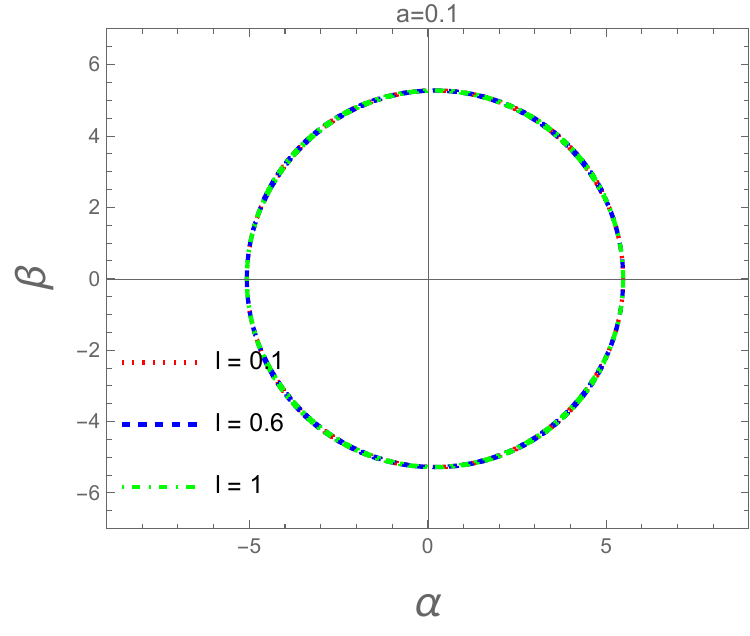}
\includegraphics[scale=0.4]{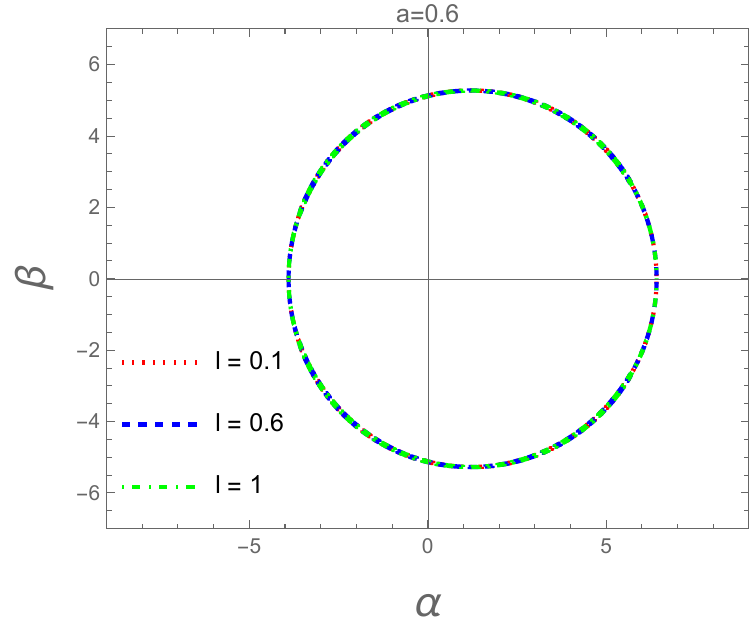}
\includegraphics[scale=0.4]{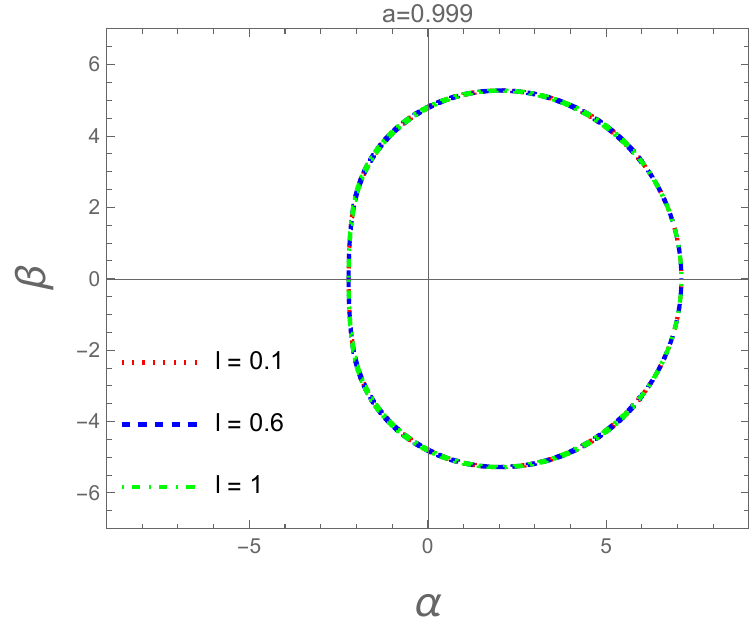}
\setlength{\abovecaptionskip}{-0.2cm}
\setlength{\belowcaptionskip}{0.3cm}
\caption{The shape of the shadow of the rotating black hole mimicker surrounded by the SC for different $\ell$ with $M=1,L=0.01$. The values of $a$ from left to right are 0.1, 0.6, 0.999, respectively. }
\label{l_vary}
\end{figure}

We first study the effect of the spin parameter $a$ on the shadow of the rotating black hole mimicker surrounded by the SC. Fig. \ref{a_l_vary} gives the shape of the shadow of the rotating black hole mimicker surrounded by the SC for different $a$ with $M=1,\ell=0.1$, where values of $L$ from left to right are 0.01, 0.05, 0.1, respectively. Fig. \ref{a_l_vary} gives the shape of the shadow of the rotating black hole mimicker surrounded by the SC for different $a$ with $M=1,L=0.01$, where values of $\ell$ from left to right are 0.1, 0.6, 1, respectively. From Fig. \ref{a_l_vary} and Fig. \ref{a_L_vary}, we can see that when the spin parameter is small, the shape of the shadow is closer to a circle with a fixed radius, but when the spin parameter is larger, the shape of the shadow is deformed. That is, the increase of the spin parameter $a$ makes the shape of the shadow present a D-shaped feature. The reason for this deformation is that in the high-speed rotating black bounces spacetime surrounded by the SC, due to the drag effect brought by its rotation, the unstable circular orbits of prograde and retrograde photons are no longer symmetrical, so that the shadow is a D-shape for the equatorial plane observer.

On the other hand, in Fig. \ref{L_vary} we study the effect of the SC parameter $L$ on the shape of the shadow of the rotating black hole mimicker surrounded by the SC, where $M=1,\ell=0.1$, and the values of $a$ from left to right are 0.1, 0.6, 0.999, respectively. One can see that when the SC parameter is gradually increased from $L=0.01$, the boundary of the black bounce shadow is continuously increasing, which is consistent for different spin parameters. But when the spin parameter $a$ is small, for example $a=0.1$, the shape of the shadow is closer to the standard circle. From the shadow of the spin parameter $a=0.999$, it can be seen that the shadow at this time becomes a D-shape again. But for a larger SC parameter $L=0.1$, the shape of the shadow is not as close to a D-shape as the shadow of $L=0.01$. Therefore, our results show that the SC have a significant impact on black bounce shadows, both in size and shape. In Fig. \ref{l_vary} we investigate the influence of the parameter $\ell$ responsible for central singularity regularization on shadow. From Fig. \ref{l_vary}, we can see that the red dotted line represents the shadow image of $\ell=0.1$, the blue dotted line represents the shadow image of $\ell=0.6$, and the green dotted line represents $\ell=1$ shaded image. One can see that an increase in $\ell$ has such a small effect on the shadow that we see that the shadows corresponding to different $\ell$ almost coincide. But when the spin $a$ is larger, the D-shaped shadow image still appears. In addition, we have verified that when $\ell$ is larger than a critical value (i.e., the spacetime line element (\ref{metric}) represents a traversable wormhole spacetime), the result shows a similar behavior to the shadow image when the spacetime line element represents a black hole spacetime. This demonstrates that we cannot distinguish whether the  rotating black hole mimicker surrounded by the SC is a black hole or a wormhole through the shadow.

\section{Radius and distortion of the shadow and energy emission rate} \label{Rs}
\begin{figure}[t!]
\includegraphics[scale=0.52]{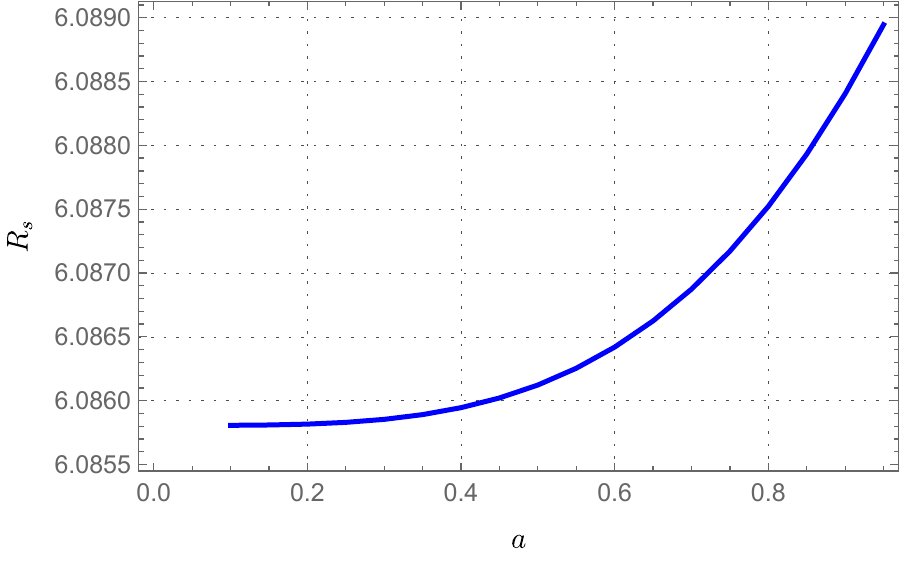}
\includegraphics[scale=0.5]{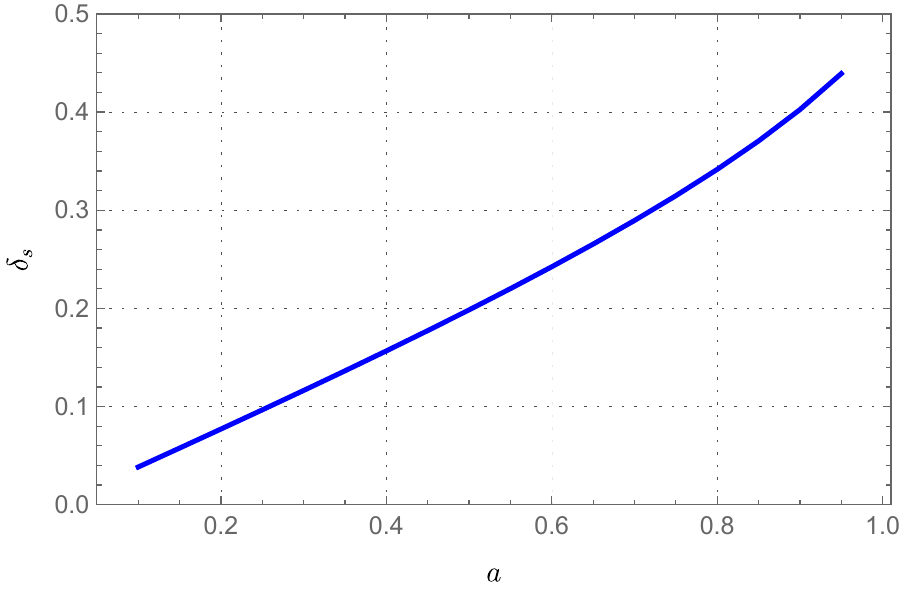}
\setlength{\abovecaptionskip}{-0.2cm}
\setlength{\belowcaptionskip}{0.9cm}
\caption{The radius (left panel) and distortion (right panel) of the shadow as the function of the spin $a$ with $M=1,\ell=0.1,L=0.1$. }
\label{Rs_a}
\includegraphics[scale=0.5]{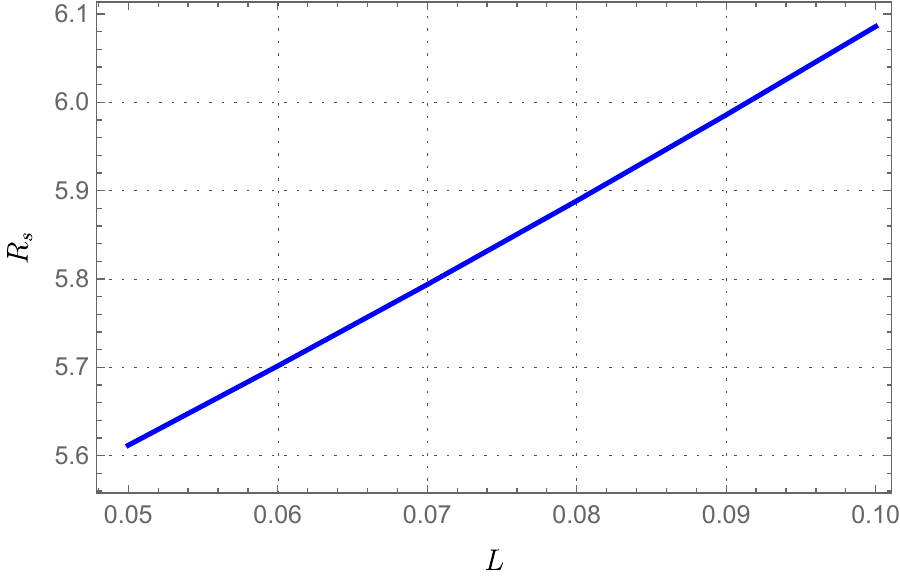}
\includegraphics[scale=0.54]{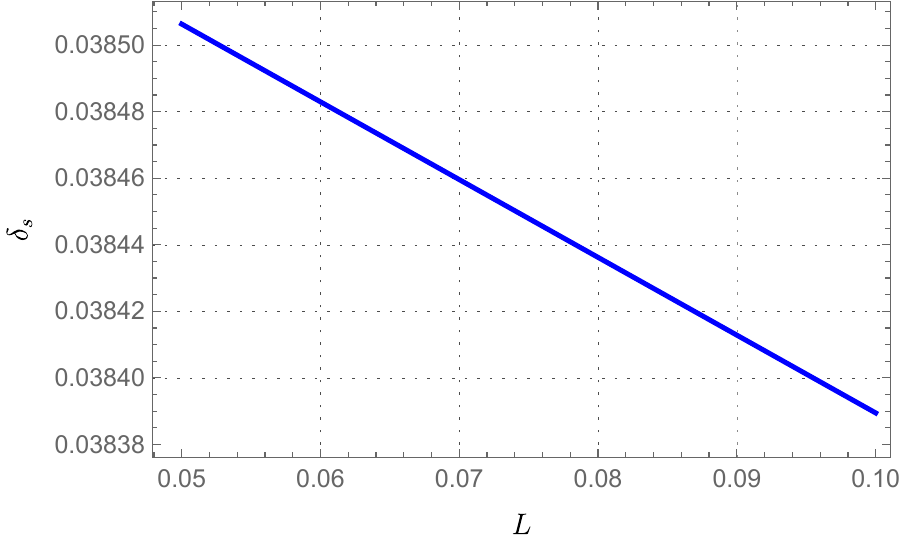}
\setlength{\abovecaptionskip}{-0.2cm}
\setlength{\belowcaptionskip}{0.9cm}
\caption{The radius (left panel) and distortion (right panel) of the shadow as the function of $L$ with $M=1,\ell=0.1,a=0.1$. }
\label{Rs-L}
\includegraphics[scale=0.5]{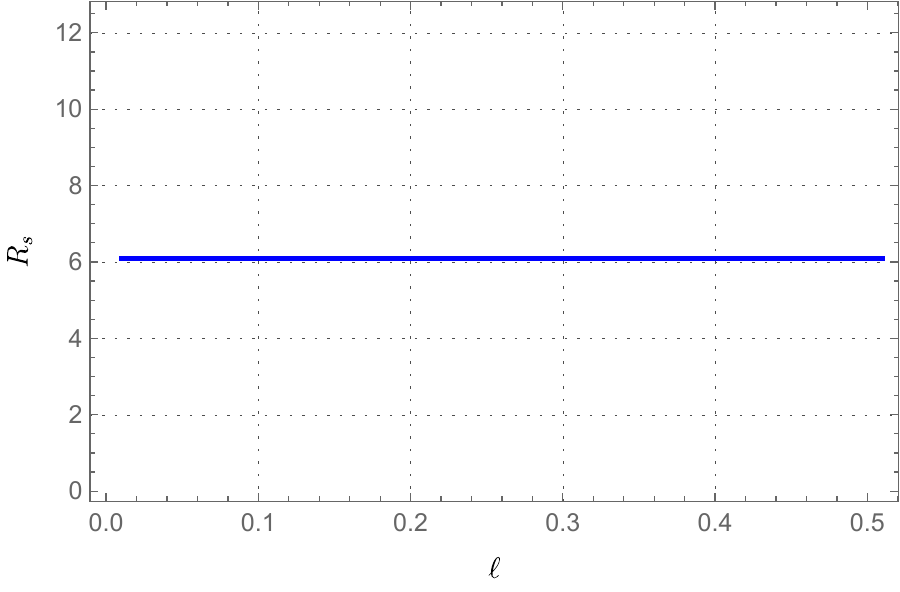}
\includegraphics[scale=0.515]{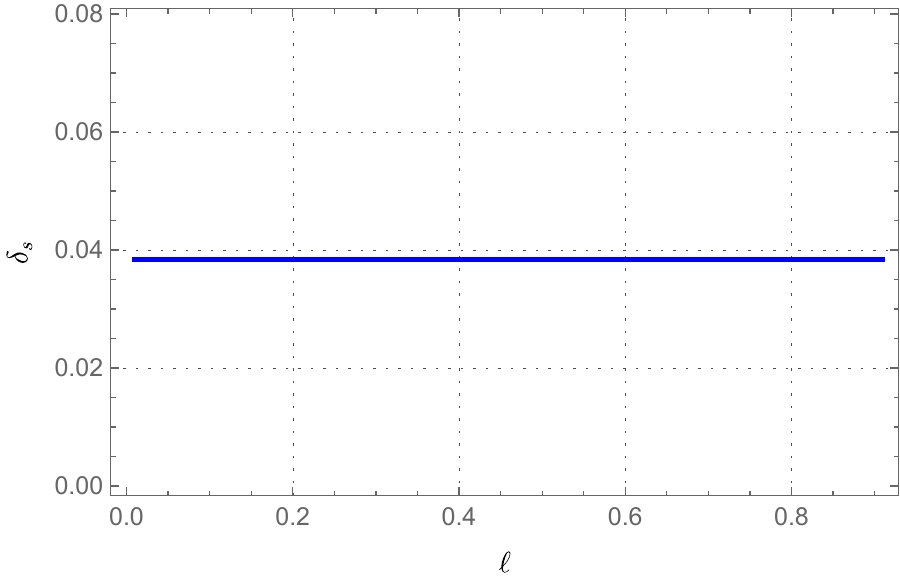}
\setlength{\abovecaptionskip}{-0.2cm}
\setlength{\belowcaptionskip}{0.8cm}
\caption{The radius (left panel) and distortion (right panel) of the shadow as the function of $\ell$ with $M=1,L=0.1,a=0.1$. }
\label{Rs_l}
\end{figure}

\begin{figure}[htbp!]
\includegraphics[scale=0.49]{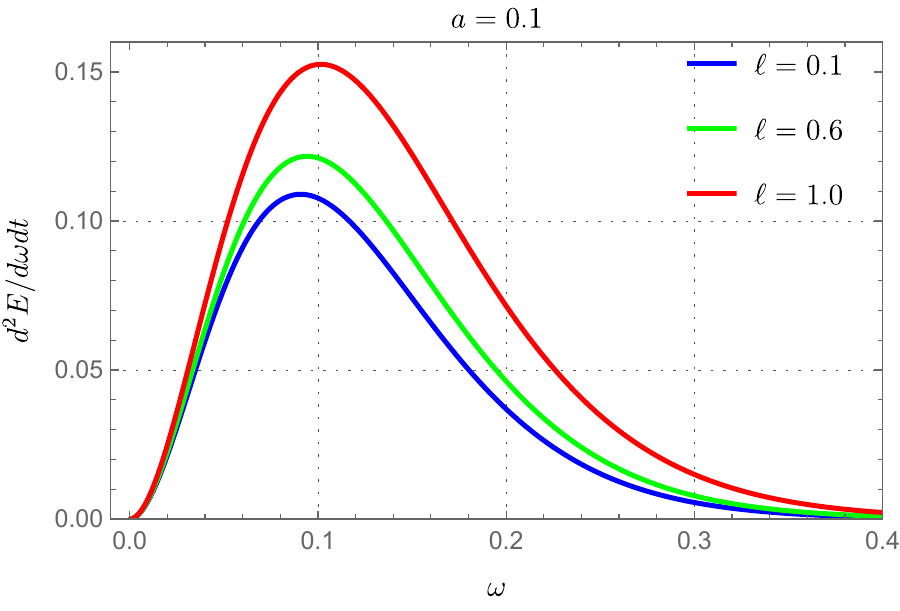}
\includegraphics[scale=0.5]{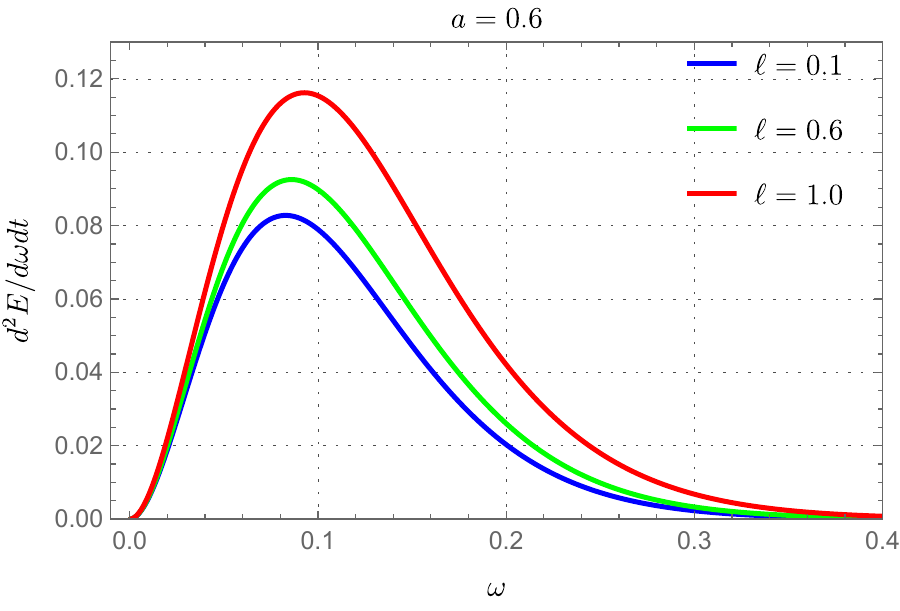}
\includegraphics[scale=0.5]{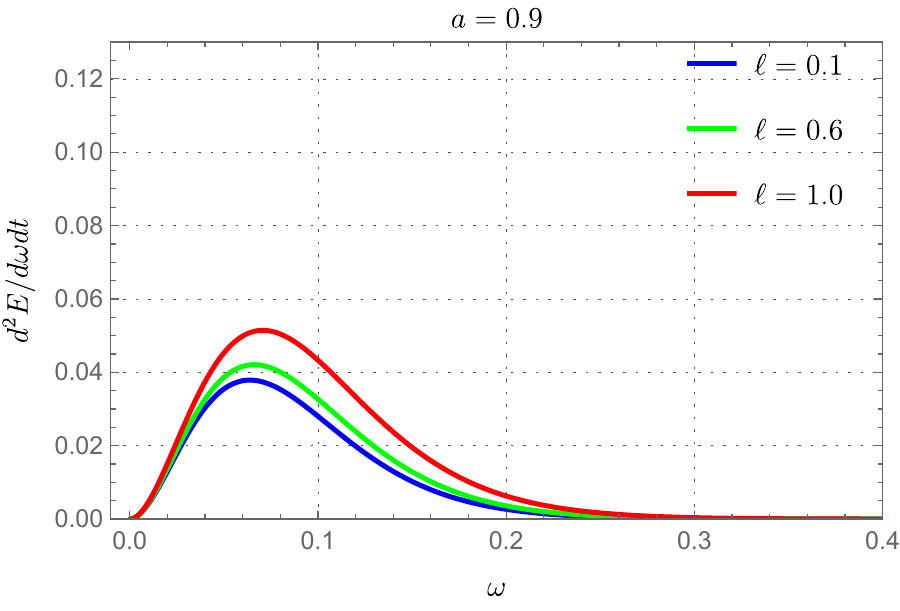}
\includegraphics[scale=0.5]{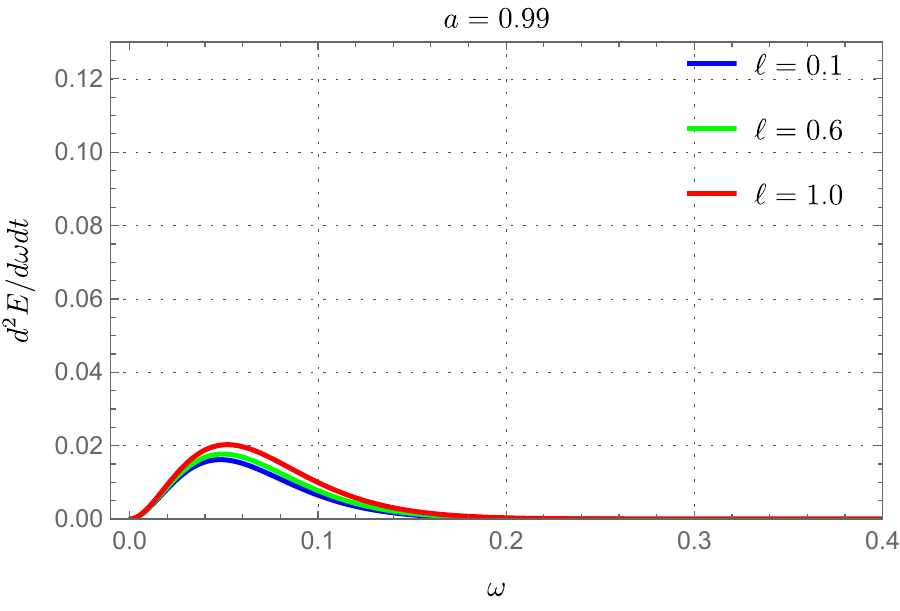}
\setlength{\abovecaptionskip}{-0.6cm}
\setlength{\belowcaptionskip}{0.5cm}
\caption{The energy emission rate as the function of the particle frequency $\omega$ for different $\ell$ with $M=1,L=0.1$. }
\label{E_lv}
\includegraphics[scale=0.49]{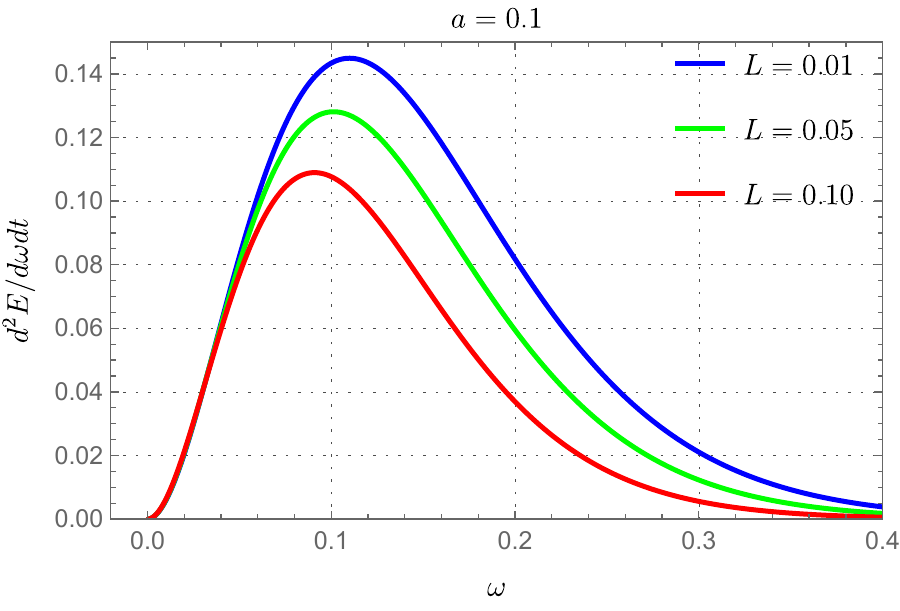}
\includegraphics[scale=0.5]{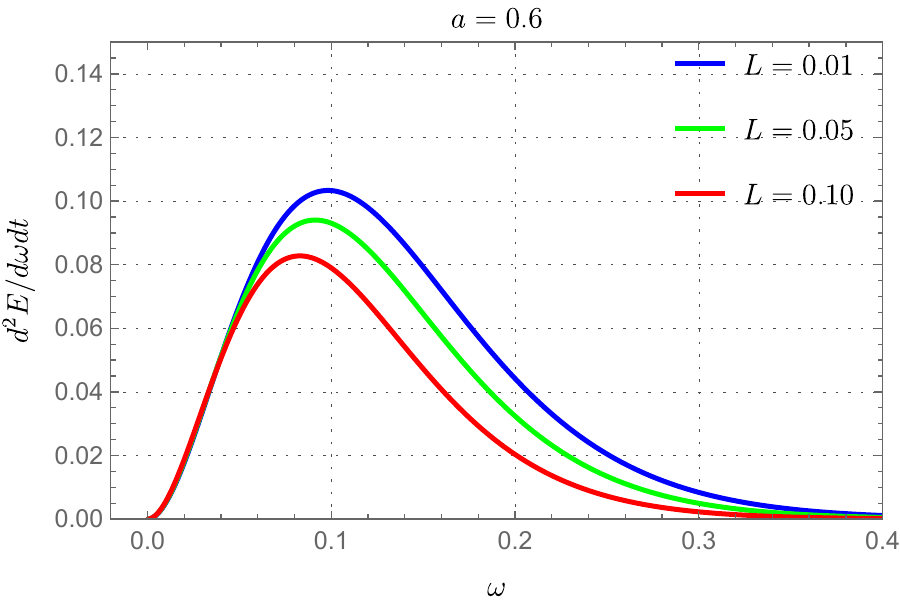}
\includegraphics[scale=0.5]{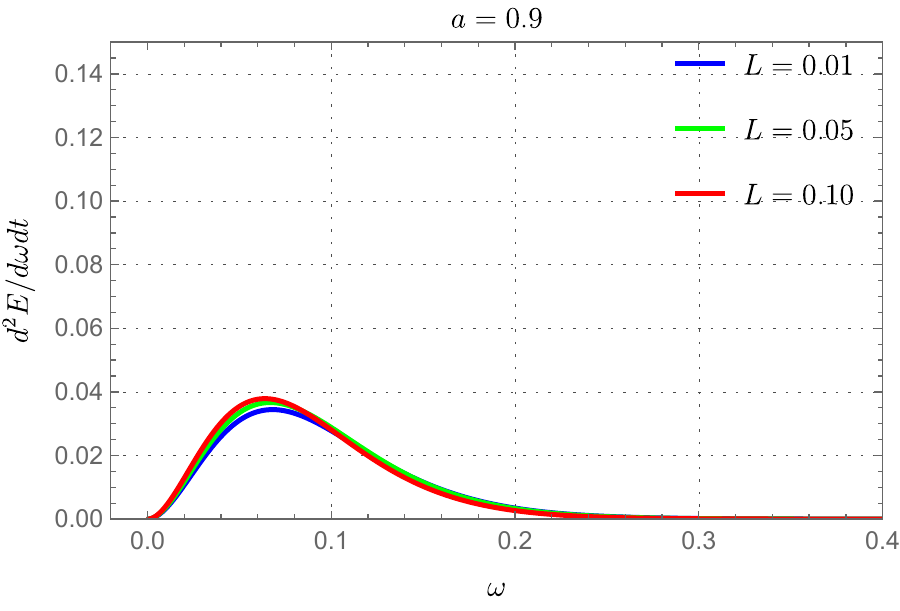}
\includegraphics[scale=0.5]{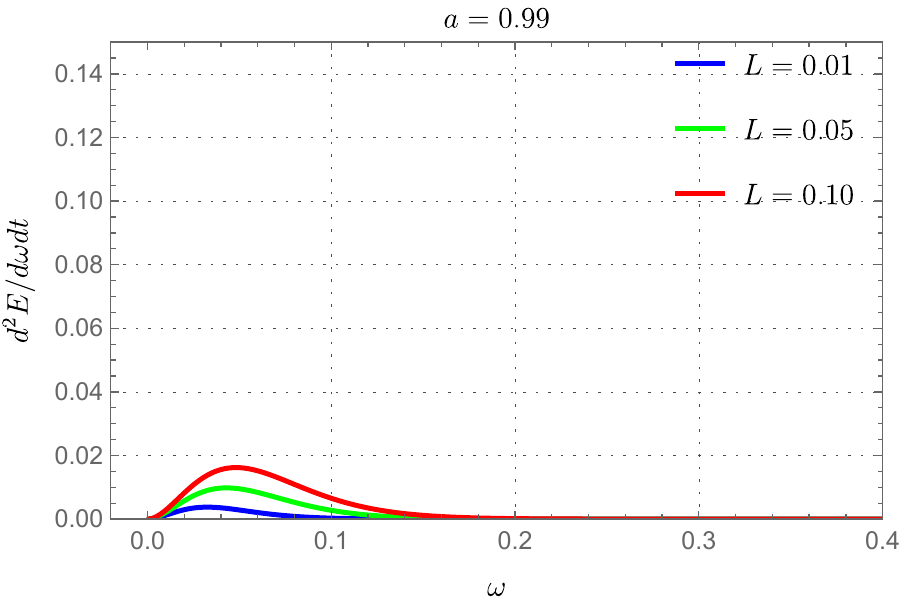}
\setlength{\abovecaptionskip}{-0.6cm}
\setlength{\belowcaptionskip}{0.3cm}
\caption{The energy emission rate as the function of the particle frequency $\omega$ for different $L$ with $M=1,\ell=0.1$. }
\label{E_Lv}
\end{figure}

The radius of the shadow and the distortion are two important observables. Here we will study the radius of the shadow and the distortion of the rotating black hole mimicker surrounded by the string cloud. The shadow radius ($R_s$) describes the scale of the black hole's shadow shape. The distortion 
 ($\delta_s$) of the shadow describes how far the boundary of the black hole's shadow deviates from the standard circle. Their specific mathematical expression is \cite{Haroon:2018ryd,Jusufi:2019nrn}
\begin{equation}
\begin{aligned}
R_s=\frac{\left(\alpha_t-\alpha_r\right)^2+\beta_t^2}{2\left|\alpha_r-\alpha_t\right|},
\end{aligned}
\end{equation}

\begin{equation}
\begin{aligned}
\delta_s=\frac{\left|\alpha_p-\tilde{\alpha}_p\right|}{R_s} .
\end{aligned}
\end{equation}
More details of radius and distortion are presented in Ref. \cite{Hioki:2009na}.
From Fig. \ref{Rs_a}, one can see that the shadow radius $R_s$ increases slowly when $a$ is small and increases sharply when $a$ is large as spin $a$ increases. We can also see that as spin $a$ increases, the distortion $\delta_s$ of rotating black hole mimicker surrounded by the string cloud increases significantly. Therefore, the shadow of the rotating black hole mimicker surrounded by the string cloud takes on a distinct D-shape when $a$ is large. 
Furthermore, from Fig. \ref{Rs-L} we can see that the radius $R_s$ of the shadow increases sharply as $L$ increases. The distortion $\delta_s$ decreases as $L$ increases. In Fig. \ref{Rs_l} we give the radius and distortion of the shadow as the function of $\ell$. 
Fig. \ref{Rs_l} shows that the shadow radius $R_s$ and shadow distortion $\delta_s$ remain almost unchanged as $\ell$ increases. That is, as $\ell$ changes, the radius of the shadow hardly changes, and there is no significant distortion.
These notable features of the rotating black hole mimicker surrounded by the string cloud's shadow radius $R_s$ and shadow distortion $\delta_s$ make it possible to be detected by EHT observations.

For an observer at infinity, the area of the black hole's shadow is very close to the high-energy absorption cross-section ($\Omega$). In a spherically symmetric black hole, the absorption section fluctuates at $\Omega$. For the shadow with the radius of $R_s$, we have $\Omega\approx \pi R_s^2$.  The energy emission rate of the black hole can be given by
\begin{equation}
\frac{d^2 E(\omega)}{d \omega d t}=\frac{2 \pi^2 \omega^3 \Omega}{e^{\frac{\omega}{T}}-1},
\end{equation}
where $\omega$ denotes the frequency of the photon and $T$ denotes the Hawking temperature at the event horizon of
the rotating black hole mimicker surrounded
by the string cloud, which can be given by \cite{Haroon:2018ryd}
\begin{equation}
T=\lim _{\theta \rightarrow 0, r \rightarrow r_{+}} \frac{1}{2 \pi \sqrt{g_{r r}}} \frac{\partial \sqrt{g_{t t}}}{\partial r} .
\end{equation}
In Fig. \ref{E_lv} and Fig. \ref{E_Lv}, we give the energy emission rate as the function of the particle frequency $\omega$. We can see that the peak of the energy emission rate gradually decreases as the spin $a$ increases. The peak of the energy emission rate increases as $\ell$ increases. Moreover, one can see that the peak of the energy emission rate decreases as $L$ increases when $a$ is small, but increases as $L$ increases when $a$ is large. This feature opens up new possibilities for the EHT to detect the rotating black hole mimicker surrounded by the string cloud.

\section{Conclusion} \label{sec:summary}
In this work, we first derive a rotating black hole mimicker surrounded by the SC using the Newman-Janis method. This rotating metric can be interpolated to represent regular black hole spacetime and wormhole spacetime. In this spacetime background, the Hamilton-Jacobi equation is solved to obtain the corresponding geodesic equation. Then we give the shadow image of this spacetime. We find that as the spin parameter $a$ increases, the shadow gradually takes on a distinct D-shape. In addition, the SC parameter $L$ has a remarkable effect on the shadow, and the increase of $L$ sharply increases the boundary of the shadow, and also makes the D-shaped shadow closer to a circle. String cloud has a very significant contribution to the shadows of the rotating black hole mimicker surrounded by the string cloud, which makes string cloud expected to be detected by EHT. Our results show that the parameter $l$ responsible for the central singularity regularization has a very weak influence on the shadow, so that we cannot use the shadow to distinguish whether the rotating black hole mimicker surrounded by the SC are the regular black holes or the wormholes. In Ref. \cite{Yang:2022ryf}, we studied the quasinormal modes of non-rotating black hole mimicker surrounded by the SC, and we found that the gravitational wave echoes appeared after the initial ringdown in this spacetime. It is very interesting to verify whether the rotating black hole mimicker surrounded by the SC will also appear gravitational wave echoes as expected in the next work. If so, the combination of shadow and gravitational wave echoes detection can help us determine whether this rotating spacetime is a black hole or a wormhole. 



\begin{acknowledgments}
This research was funded by the National Natural Science Foundation of China (No.12265007 and No.12261018), Universities Key Laboratory of System Modeling and Data Mining  in Guizhou Province (No.2023013), the Science and Technology Foundation of Guizhou Province (No. ZK[2022]YB029). The work of G.L.  is supported by the Italian Istituto Nazionale di Fisica Nucleare (INFN) through the ``QGSKY'' project and by Ministero dell'Istruzione, Universit\`a e Ricerca (MIUR). G.L.,  and A. {\"O}.  would like to acknowledge networking support by the COST Action CA18108 - Quantum gravity phenomenology in the multi-messenger approach (QG-MM). A. {\"O}. would like to acknowledge networking support by the COST Action CA21106 - COSMIC WISPers in the Dark Universe: Theory, astrophysics and experiments (CosmicWISPers).
\end{acknowledgments}

\appendix
\section{String cloud model}\label{Appendix}

In this Appendix, we review the main features of spacetimes in the presence of a cloud of strings, working in the framework of General Relativity. 
The basic idea underlying the string model is that a string is associated with a world sheet (in analogy to a particle which is associated with a world line), which is described by $x^\mu(\lambda^A)$, where $\lambda^0$ and $\lambda^1$ are timelike and spacelike parameters.
General Relativity minimally coupled with matter and the cloud of strings is described by the total action  
\begin{equation}
    S= S_{GR}+S_M+S_{CS},\label{Actiongeral}
\end{equation}
where $S_{GR}=\int d^4x \sqrt{-g}R$ is the Hilbert-Einstein action, $S_M$ is the matter action, and $S_{CS}$ is the Nambu--Goto action that describes string-like objects, whose expression is given by \cite{Letelier:1979ej}
\begin{equation}\label{ActionCS1}
    S_{CS}= \mathcal{M} \, \int \sqrt{-\gamma}d\lambda^0d\lambda^1.
\end{equation}
In (\ref{ActionCS1}), $\mathcal{M}$ is a dimensionless constant characterizing the string, while $\gamma=\det \gamma_{AB}$, with $\gamma_{AB}$ the induced metric on a sub-manifold defined as
\begin{equation}
    \gamma_{AB}=g_{\mu\nu}\frac{\partial x^\mu}{\partial \lambda^A}\frac{\partial x^\nu}{\partial \lambda^B}\,,
    \qquad \lambda^A= 0, 1\,.
\end{equation}
The Nambu--Goto action can be rewritten as
\begin{equation}
    S_{CS}=\int \mathcal{M}\left(-\frac{1}{2}\Sigma^{\mu\nu}\Sigma_{\mu\nu}\right)^{1/2}d\lambda^0d\lambda^1,
\end{equation}
where $\Sigma^{\mu\nu}$ is the bi-vector defined as
\begin{equation}\label{SigmaG}
    \Sigma^{\mu\nu}=\epsilon^{AB}\frac{\partial x^\mu}{\partial \lambda^A}\frac{\partial x^\nu}{\partial \lambda^B},
\end{equation}
being $\epsilon^{AB}$ the Levi--Civita symbol with non-vanishing components given by $\epsilon^{01}=-\epsilon^{10}=1$. The variation of the action (\ref{Actiongeral}) with respect to the metric tensor gives the field equations (\ref{einstein}), where
%
$T^{M}_{\mu\nu}$ is the usual stress-energy tensor of the matter, and $T^{CS}_{\mu\nu}$ the stress-energy tensor of the string cloud, given by (\ref{TensorCSG}) \cite{Letelier:1979ej}.
Moreover, the stress-energy tensor $T^{CS}_{\mu\nu}$ satisfies the conservation law
\begin{equation}
    \nabla_\mu {T^{CS}}^{\mu\nu}=\nabla_\mu\left(\frac{\rho\Sigma^{\mu \alpha}{\Sigma_{\alpha}}^\nu}{8\pi\sqrt{-\gamma}}\right)=\nabla_\mu\left(\rho\Sigma^{\mu \alpha}\right)\frac{{\Sigma_{\alpha}}^\nu}{8\pi\sqrt{-\gamma}}+\rho\Sigma^{\mu \alpha}\nabla_\mu\left(\frac{{\Sigma_{\alpha}}^\nu}{8\pi\sqrt{-\gamma}}\right)=0.
\end{equation}
{\bf Simpson--Visser solution with the cloud of strings -} The standard Simpson--Visser (SV) solution is characterized by the fact that it describes a regular black hole (and, depending on the parameters of the metric, it can describe a wormhole). To obtain the SV solution in the presence of a cloud of string, the energy-momentum tensor ${T_{\mu\nu}}^M$ in (\ref{einstein}) is given by (\ref{TensorMG}).
The non--vanishing components of the stress-energy tensor $T_{\mu\nu}^{SV}$, associated with the SV solution, are \cite{Simpson:2018tsi}
\begin{eqnarray}
&&T^{SV}_{00}=-\frac{l^2 \left(\sqrt{l^2+r^2}-4 M\right) \left(\sqrt{l^2+r^2}-2 M\right)}{8\pi\left(l^2+r^2\right)^3}, T^{SV}_{11}=\frac{l^2}{8\pi\left(l^2+r^2\right)^{3/2} \left(2 M-\sqrt{l^2+r^2}\right)},\label{TSV}\\
&&T^{SV}_{22}=\frac{l^2 \left(\sqrt{l^2+r^2}-M\right)}{8\pi\left(l^2+r^2\right)^{3/2}},\qquad T^{SV}_{33}=\frac{l^2 \sin ^2\theta  \left(\sqrt{l^2+r^2}-M\right)}{8\pi\left(l^2+r^2\right)^{3/2}}\,. \nonumber
\end{eqnarray}
The non--vanishing components of the stress-energy tensor $T^{NMC}_{\mu\nu}$, which provides information about the non-minimal coupling between the SV metric and the cloud of string, are
\begin{eqnarray}
   && T^{NMC}_{00}=-\frac{L l^2 \left((L-2) \sqrt{l^2+r^2}+6 M\right)}{8\pi\left(l^2+r^2\right)^{5/2}},\nonumber\\
    &&T^{NMC}_{11}=\frac{2 L l^2 M}{8\pi\left(l^2+r^2\right)^{3/2} \left(\sqrt{l^2+r^2}-2 M\right) \left((L-1) \sqrt{l^2+r^2}+2 M\right)},\label{TNMC}\\
    &&T^{NMC}_{22}=-\frac{L l^2}{8\pi\left(l^2+r^2\right)},\qquad T^{CP}_{33}=-\frac{L l^2 \sin ^2\theta }{8\pi\left(l^2+r^2\right)}\nonumber.
\end{eqnarray}
Notice that for $L=0$ or $l=0$ all the components of $T^{NMC}_{\mu\nu}$ vanish.

Finally, the non-vanishing components of the cloud of strings $T^{CS}_{\mu\nu}$ are
\begin{equation}\label{TCS}
    {{T^{CS}}^{0}}_0={{T^{CS}}^{1}}_1=\frac{L}{8\pi \left(r^2+l^2\right)}.
\end{equation}
To guarantee the positivity of the energy density it is required $L>0$.
By using Eqs. (\ref{einstein}) and (\ref{TSV}, \ref{TNMC}, \ref{TCS}) one infers a set of equations leading to metric function (\ref{huar}).

\bibliography{ref}
\bibliographystyle{AAA}
\end{document}